\documentclass[conference]{IEEEtran}
\IEEEoverridecommandlockouts
\usepackage{cite}
\usepackage{amsmath,amssymb,amsfonts}
\usepackage{algorithmic}
\usepackage{graphicx}
\usepackage{textcomp}
\usepackage{xcolor}
\usepackage{hyperref}
\usepackage{cite}
\usepackage{amsmath,amssymb,amsfonts}
\usepackage{algorithmic}
\usepackage{graphicx}
\usepackage{textcomp}
\usepackage{xcolor}
\usepackage{subfigure}
\usepackage{caption}
\usepackage{subcaption}
\usepackage{float}

\def\mB{\mbox{$\mathbf{B}$}}

\def\mI{\mbox{$\mathbf{I}$}}
\def\mL{\mbox{$\mathbf{L}$}}
\def\mY{\mbox{$\mathbf{Y}$}}

\def\mS{\mbox{$\mathbf{S}$}}

\def\mU{\mbox{$\mathbf{U}$}}

\def\mq{\mbox{$\mathbf{q}$}}
\def\ms{\mbox{$\mathbf{s}$}}
\def\my{\mbox{$\mathbf{y}$}}
\newcommand{\ds}{\displaystyle}

\newcommand{\beq}{\begin{equation}}
\newcommand{\eeq}{\end{equation}}
\usepackage{float}

\def\BibTeX{{\rm B\kern-.05em{\sc i\kern-.025em b}\kern-.08em
    T\kern-.1667em\lower.7ex\hbox{E}\kern-.125emX}}
\begin{document}
\title{Learning Higher-Order Interactions in Brain Networks via Topological Signal Processing\\
{\footnotesize 
\thanks{This work was funded by the FIN-RIC Project TSP-ARK, supported by Universitas Mercatorum under grant No. 20-FIN/RIC. Additional support was provided by CAPES (88881.311848/2018-01, 88887.899136/2023-00), CNPq (442238/2023-1, 312935/2023-4, 405903/2023-5), and FACEPE (APQ-1226-3.04/22).
}
}}

\author{\IEEEauthorblockN{Breno C. Bispo$^1$, Stefania Sardellitti$^2$, Fernando A. N. Santos$^3$, Juliano B. Lima$^1$}
\IEEEauthorblockA{$^1$Dept. of Electronics and Systems, Federal University of Pernambuco, Recife, Brazil\\
$^2$Engineering and Sciences Dept., Universitas Mercatorum, Rome, Italy
\\
$^3$Dutch Institute for Emergent Phenomena, KdVI, University of Amsterdam,  Amsterdam, The Netherlands
}
e-mail: breno.bispo@ufpe.br, stefania.sardellitti@unimercatorum.it, f.a.nobregasantos@uva.nl, juliano.lima@ufpe.br
}

\maketitle


\begin{abstract}

Our goal in this paper is to leverage the potential of the topological signal processing (TSP) framework for analyzing brain networks. Representing brain data as signals over simplicial complexes allows us to capture higher-order relationships within brain regions of interest (ROIs).
Here, we focus on learning the underlying brain topology from observed neural signals  using  two distinct inference strategies. The first method relies on  higher-order statistical metrics to infer multiway relationships among ROIs. The second method jointly learns the brain topology and  sparse signal representations, of both the solenoidal and harmonic components of the signals, by minimizing the total variation along triangles and the data-fitting errors. 
Leveraging the properties of solenoidal and irrotational signals,  and their physical interpretations, we extract functional connectivity features from brain topologies and uncover new insights into functional organization patterns. 
This allows us to associate brain functional connectivity (FC) patterns of conservative signals with well-known functional segregation and integration properties. Our findings align with recent neuroscience research, suggesting that our approach may offer a promising pathway for characterizing the higher-order brain functional connectivities.
\end{abstract}


\section{Introduction}

Advances in neuroimaging technologies, particularly functional magnetic resonance imaging (fMRI), have enabled non-invasive, high-resolution mapping of brain activity. 
This has allowed for indirect explorations of anatomical and functional interactions between distinct brain regions of interest (ROIs), improving our understanding of neural activity and facilitating the development of  predictive models of brain function. 
Modeling brain networks as graphs has become a natural approach for representing these complex systems, where nodes correspond to ROIs and links capture functional and structural connectivity~\cite{bassett2017network}. 
Recently, graph signal processing (GSP) tools, 
which extend classical discrete-time signal processing concepts to signals defined on graphs, have been extensively employed to analyze neural signals observed at the nodes of the brain network~\cite{li}. 
However, graph-based models are limited to capturing only dyadic relationships within the data, restricting their ability to fully characterize higher-order structures and functions. 
On the other hand, recent studies~\cite{varley, rosas} have leveraged multivariate interdependence metrics to infer higher-order interactions (HOIs) among ROIs, i.e. simultaneous inter-dependence among three or more nodes, aiming to uncover novel FC patterns that remain elusive to solely pairwise-based approaches. 
To efficiently characterize HOIs,  brain networks have been represented on more structured topological spaces than graphs, such as hypergraphs~\cite{battiston}. 

More recently, hypergraph signal processing (HGSP) has shown potential applications in characterizing higher-order FC~\cite{breno}. However, the interpretation of its algebraic operators remains a challenge in extracting meaningful insights into the dynamics of HOIs. Nevertheless, a particular class of hypergraphs, such as simplicial complexes, offers a rich algebraic structure enhancing data interpretability and features extraction. 
Built upon this algebraic structure, a recently developed framework Topological Signal Processing (TSP)~\cite{barb_2020} has emerged as a powerful tool for  analyzing and processing  signals defined over simplicial complexes. This framework promises to enhance brain data analysis and interpretability by  enabling the  identification of critical cycles and detection of brain regions with significant activity.

One of the main challenges in processing information over simplicial complexes is learning their underlying structures from data in a way that accurately reflects their features. In this paper, we address this challenge by developing strategies to learn the brain simplicial complex (SCX) topology from real brain datasets using two different approaches. The first approach leverages (non-)dyadic statistical metrics to infer the presence of edges and filled triangles, capturing pairwise and triple-wise interdependencies between ROIs of the brain SCX. The second approach aims to jointly determine the brain SCX topology and sparse signal representations by minimizing the data-fitting error and the total variation of the signals.  Analyzing HOIs in brain networks through the lens of TSP, we leverage key operators that provide physical interpretations of signals over a SCX, such as the divergence and curl operators. We analyze resting-state fMRI (rs-fMRI) signals to identify the primary sources and sinks of information in the brain, as well as significant circulations among ROIs within our learned topologies. Our findings reveal that the mean divergence patterns in the brain align with established neuroscience literature. Additionally, the observed conservative circulations remind the idea of functional segregation and integration, reflecting key organizational principles of brain functioning. 

\section{Overview of Topological Signal Processing}\label{sec:overview}
In this section we briefly recall the basic TSP principles \cite{barb_2020}.
Let us consider a  finite set of  $N$ nodes $\mathcal{V}=\{v_0, v_1, \cdots, v_{N-1}\}$. A $k$-simplex $\tau^k=\{ v_{j_0},\ldots, v_{j_k}\}$ is an unordered subset of $\mathcal{V}$ with $k+1$ distinct nodes. A face of the $k$-simplex $\tau^k=\{ v_{j_0},\ldots, v_{j_k}\}$ is a $(k-1)$-simplex $\tau^{k-1}=\{ v_{j_0},\ldots, v_{j_{n-1}},v_{j_{n+1}}, \ldots v_{j_k}\}$ for some $0\leq n \leq k$.  
Therefore, a node is a 0-simplex, an edge  a 1-simplex, a triangle  a 2-simplex,  and so on. 
A simplicial complex (SCX) $\mathcal{X}^K$ of order $K$ is a finite set of $k$-simplices $\tau^k$ for $k\in\{0,1,\cdots, K\}$, that is closed under the inclusion property, i.e. if $\tau_i^k \in \mathcal{X}^K$, all faces of $\tau_i^k$ are in $\mathcal{X}^K$. 
The order of a simplicial complex  is the maximum dimension of its simplices.  
The topological structure of a simplicial complex is captured by the incidence relations between its simplices.
Specifically, two simplices of order $k$, $\tau_i^k$ and
$\tau_j^k$ are upper adjacenct in $\mathcal{X}$ if both are faces 
of a simplex of order $k+1$ or
lower adjacent if they have a common face of order $k$. A $(k-1)$-face $\tau_j^{k-1}$ of a $k$-order simplex $\tau_i^{k}$ is called a boundary element of $\tau_i^{k}$, denoted as $\tau_j^{k-1} \prec_b \tau_i^{k}$.
Given an orientation of the simplices \cite{barb_2020}, we denote as $\tau_i^{k-1} \sim \tau_j^{k}$  two  coherently oriented simplices. Then, the structures of the simplicial complex is described by the incidence matrices   
$\mathbf{B}_k$, whose entries establish which $(k-1)$-simplices are  incident to which $k$-simplices.

Specifically, we have $B_k(i,j)=1$ (or $B_k(i,j)=-1$) if $\tau_i^{k-1} \prec_b \tau_j^{k}$ and  $\tau_i^{k-1} \sim \tau_j^{k}$ (or $\tau_i^{k-1} \not\sim \tau_j^{k}$), while $B_k(i,j)=0$ if $\tau_i^{k-1} \not\prec_b \tau_j^{k}$. An important property of the incidence matrices is that it always holds $\mB_k\mB_{k+1}=\mathbf{0}$.
In this work, we restrict our attention w.l.o.g. to simplicial complexes $\mathcal{X}^2$ of  order $K=2$ due to practical application and computational feasibility.  Let us denote with $\mathcal{V}$, $\mathcal{E}$ and $\mathcal{T}$, the set of vertices, edges and triangles, respectively, with cardinalities 
$|\mathcal{V}|=N$, $|\mathcal{E}|=E$, $|\mathcal{T}|=T$.
Given the node-edge incidence matrix $\mathbf{B}_1\in\mathbb{R}^{N\times E}$ and the edge-triangles incidence matrix $\mathbf{B}_2\in\mathbb{R}^{E\times T}$, the topological structure of $\mathcal{X}^2$ is described by  the graph Laplacian $\mathbf{L}_0=\mathbf{B}_1\mathbf{B}_1^T\in\mathbb{R}^{N\times N}$, which encodes the upper adjacencies of 0-simplices, and  by the first-order Hodge Laplacian matrix \cite{goldberg2002combinatorial}
$\mathbf{L}_1=\mathbf{B}_1^T\mathbf{B}_1 + \mathbf{B}_{2}\mathbf{B}_{2}^T\in\mathbb{R}^{E\times E},$
where $\mathbf{L}_{1,\ell}=\mathbf{B}_1^T\mathbf{B}_1$
and $\mathbf{L}_{1,u}=\mathbf{B}_{2}\mathbf{B}_{2}^T$ are the lower and upper Laplacian matrices, respectively. They encode the lower and upper adjacencies of all $1$-simplices, respectively~\cite{barb_2020}.\\
\textbf{Topological signals representation}. Denoting by $N_k$ the number of $k$-order simplices, a $k$-simplicial signal $\mathbf{s}^k=[s^k(1),\cdots, s^k(N_k)]^T\in\mathbb{R}^{N_k}$ is a map $\mathbf{s}^k:\mathcal{S}^k\rightarrow \mathbb{R}^{N_k}$, where the entry $s^k(i)$ corresponds to the $i$-th $k$-simplex~\cite{barb_2020} and $\mathcal{S}^k$ is the subset of simplices of order $k$.
We denote with  $\mathbf{s}^0\in\mathbb{R}^{N},\mathbf{s}^1\in\mathbb{R}^{E},\mathbf{s}^2\in\mathbb{R}^{T}$ the  signals over the $N$ nodes, $E$ edges and $T$ triangles of $\mathcal{X}^2$, respectively.
Suitable bases to represent $k$-simplicial signals  are the eigenvectors of the $k$-th order Hodge Laplacian matrix \cite{barb_2020}. Let us consider the eigendecomposition $
\mathbf{L}_k=\mathbf{U}_k\mathbf{\Lambda}_k\mathbf{U}^{T}_k$ with 
$\mathbf{U}_{k}\in\mathbb{R}^{N_k\times N_k}$ the eigenvectors matrix and $\mathbf{\Lambda}_{k}$  the diagonal matrix with entries the associated eigenvalues $\lambda_{k,i}$, $i=1,\ldots,N_k$. The Simplicial Fourier Transform (SFT) \cite{barb_2020} of $\ms^{k}$ can be defined as  $\hat{\ms}^{k}=\mU_k^{T}\ms^k$ and, then, the signal can be represented in terms of its SFT coefficients as $\ms^k=\mU_k \hat{\ms}^{k}$. A signal $\ms^k$ is bandlimited  if  it  admits a sparse representation on the eigenvectors bases.

One of the key aspects in TSP that bridges mathematical concepts to physical interpretations is the possibility to decompose the space of a $k$-simplicial signal $\mathbb{R}^{N_k}$ into the direct sum of three orthogonal subspaces according to the so called Hodge decomposition 
\vspace{-0.2cm}
\begin{equation}\label{eq:hodge_decomposition}
    \mathbb{R}^{N_k} \equiv \text{img}(\mathbf{B}_k^T) \oplus \text{ker}(\mathbf{L}_k) \oplus \text{img}(\mathbf{B}_{k+1})\:.
\end{equation}
In the domain of the edge flows (1-simplices), which can be considered as a discrete counterpart of vector fields~\cite{barb_2020}, $\text{img}(\mathbf{B}_1^T)$, $\text{img}(\mathbf{B}_{2})$ and $\text{ker}(\mathbf{L}_1)$ are the gradient, curl and harmonic subspaces with dimensions $N_G$, $N_C$ and $N_H$, respectively. Therefore, we can partition the  eigenvectors of $\mathbf{U}_1$ into three groups of eigenvectors:
    i) the gradient eigenvectors, $\mathbf{U}_{\text{G}}\in\mathbb{R}^{E\times N_G}$, consisting of eigenvectors of $\mathbf{L}_{1,\ell}$ 
    (associated with non-zeros eigenvalues) 
    that generates the gradient subspace $\text{img}(\mathbf{B}_1^T)$;
    ii) the curl (or solenoidal) eigenvectors $\mathbf{U}_{\text{C}}\in\mathbb{R}^{E\times N_C}$,  i.e. the  eigenvectors of $\mathbf{L}_{1,u}$, associated with non-zeros  eigenvalues and 
spanning the curl subspace $\text{img}(\mathbf{B}_2)$; and iii)
    the harmonic eigenvectors $\mathbf{U}_{\text{H}}\in\mathbb{R}^{E\times N_H}$  spanning the kernel of $\mathbf{L}_{1}$. The dimension of $\text{ker}(\mL_1)$, known as the Betti number of order $1$, represents a key  topological invariant of the SCXs, counting the number of holes (unfilled triangles) within the complex.


From (\ref{eq:hodge_decomposition}), each  edge signal observed over the SCX admits the following Hodge decomposition
\vspace{-0.1cm}
\begin{equation} \label{eq:Hodge}
    \ms^1 =\mB_1^T \ms^0+\mB_2 \ms^2 +\ms_H\:,
\end{equation}
with $\ms_{irr}^1=\mB_1^T \ms^0$ being
the irrotational edge signal, i.e. the gradient of the node signal observed at the vertices of each edge, $\ms_{s}^1=\mB_2 \ms^2$ the solenoidal signal reflecting the circulation (curl) along triangles, and, finally, $\ms_H$ the harmonic signal component.
Two key operators that provide meaningful physical interpretations related to the signals observed over a SCX are the divergence and curl operators~\cite{barb_2020}. 
The \textit{divergence operator}, defined as
\vspace{-0.3cm}
\begin{equation}\label{eq:div_operator}
    \text{div}(\mathbf{s}^1) = \mathbf{B}_1 \mathbf{s}^1\:,
\end{equation}
\noindent is a node signal, revealing which nodes act as sources or sinks of information  within a SCX.  The \textit{curl operator}, given by  
\vspace{-0.1cm}
\begin{equation}\label{eq:curl_operator}
    \text{curl}(\mathbf{s}^1) = \mathbf{B}_2^T \mathbf{s}^1\:,
\end{equation}  
\noindent quantifies the circulation of flows along the triangles. Note that solenoidal signals  are divergence-free, since it holds $\mB_1\ms_{s}^1=\mathbf{0}$, while irrotational signals  are curl-free since we have $\mB_2^T\ms_{irr}^1=\mathbf{0}$.

\section{Statistical Learning of  the brain topology}\label{sec:statistical_learning}


In this section, we propose a method based on statistical metrics to infer a 2-order brain SCX from real resting-state fMRI (rs-fMRI)  signals, where each rs-fMRI time series, in a brain dataset with $N$ ROIs, is treated as a node signal. To construct the skeleton of a brain SCX (i.e., the graph) from node signals, we leverage a conventional pairwise-based metric commonly used in neuroscience, the Pearson correlation. 
Among the various multivariate interaction metrics used in neuroscience to infer HOIs, including multivariate cumulants~\cite{rikkert}, time series co-fluctuations~\cite{santoro2024higher}, and multivariate information-theoretic measures~\cite{timme, rosas}, we draw inspiration from the latter. These information-theoretic approaches have been shown to reveal unique higher-order organizational patterns in the brain that may remain elusive by solely pairwise-based methods~\cite{qiang, santos_2023}. Specifically, within our context, the most suitable metric for determining the weights of 2-simplices (triangles), representing the strength of three-way simultaneous interdependence, seems to be the \textit{total correlation} (TC), defined as~\cite{timme}
\begin{equation}\label{eq:tc}            \text{TC}_{\mathbf{x}_i\mathbf{x}_j\mathbf{x}_k}=\text{H}(\mathbf{x}_i)+\text{H}(\mathbf{x}_j)+\text{H}(\mathbf{x}_k)-\text{H}(\mathbf{x}_i,\mathbf{x}_j,\mathbf{x}_k)
        \end{equation}
\noindent where $\mathbf{x}_i$, $\mathbf{x}_j$ and $\mathbf{x}_k$ are random variables vectors. In our context, these random vectors represent rs-fMRI time series of ROIs observed over the distinct graph vertices  $v_i,v_j,v_k$ for $i, j, k\in\{1,\ldots,N\}$, $\text{H}(\:\cdot\:)$ is the Shannon entropy, while $\text{H}(\:\cdot\:,\:\cdot\:)$ and $\text{H}(\:\cdot\:,\:\cdot\:,\:\cdot\:)$ are the joint-entropies of the arguments. This approach is justified by the complementary property shared between (non-)pairwise correlations that characterize redundant-dominated structures~\cite{timme}. Applying this metric to a set of $N$ rs-fMRI time series, we can compute all HOI weights. 

Using the Human Connectome Project (HCP) public dataset (\href{http://www.humanconnectome.org/}{http://www.humanconnectome.org/}), we select 226 sets of $N=116$ rs-fMRI time series obtained from a cohort of 126 unrelated healthy individuals (young adults aged 21-35 consisting of 60 males and 66 females). Following the Schaefer's brain atlas reference used in Fig. 1(a) of \cite{breno}, each ROI is mapped to its corresponding functional subnetwork (FS) and color-coded accordingly.

Our goal is to study the 2-order brain SCX that best corresponds to the mean individual at HCP dataset. To achieve this, we first compute the absolute values of the Pearson correlations from each volunteer's rs-fMRI time series, represented as entries in a symmetric correlation matrix available at \href{https://doi.org/10.5281/zenodo.6770120}{(https://doi.org/10.5281/zenodo.6770120)}. We then calculate their average and retain the top 5\% strongest edge connections, which results in 333 edges. While no consensus exists on the best metric for inferring pairwise relationships, some suggest that anticorrelations hold biological significance. Thus, negative Pearson correlation values can be converted to absolute values to preserve potential insights. 

Subsequently, we compute all $253460$ triangle weights for each set of rs-fMRI time series using \eqref{eq:tc}. To enable population-level comparisons, we compute the \emph{z-score} of all triangle weights at the individual level and then take their average. From the mean individual triangle weights, we select the 200 strongest HOIs as filled triangles in the mean brain SCX. 
To maintain the inclusion property of our brain SCX, we incorporate edges that are not subsets of the filled triangles. This results in a learned 2-order brain SCX with \( N = 116 \) nodes, \( E = 491 \) edges, and \( T = 200 \) filled triangles. Furthermore, after selecting  orientations for simplices, we build the incidence matrices $\mB_1$ and $\mB_2$. It is important to emphasize that the orientation of edges and triangles for defining the incidence matrices  is merely a formalism and does not imply causality in information exchange between brain ROIs, as we are considering undirected SCXs.
In order to provide an overall analysis of the brain signals over the learned mean brain SCX, we first compute the mean individual node signals by z-scoring the rs-fMRI time series at the individual level and then averaging them across all volunteers' time series, resulting in a set of mean node signals given by the matrix $\mathbf{S}^0=[\mathbf{s}^0(1),\cdots,\mathbf{s}^0(M)] \in \mathbb{R}^{N\times M}$, where $M=2400$ is the number of time samples of the entire rs-fMRI recording. Given the node  signals, we derive the edge signals by computing the 1-order instantaneous co-fluctuation magnitude. This is obtained by taking the absolute value of element-wise multiplication of pairs of z-scored node signals, a common approach in neuroscience for assessing topological organizations from fMRI signals~\cite{santoro2024higher}. 
As a result, the edge signals are represented by the matrix \(\mathbf{S}^1=[\mathbf{s}^1(0),\dots,\mathbf{s}^1(M)]\in \mathbb{R}^{E\times M}\). 

\begin{figure}[!htbp]
    \centering
    \subfigure[Histogram of $\overline{\text{div}}(\mathbf{S}^1)$]
    {
    \includegraphics[width=6.5cm,height=2.4cm]
    {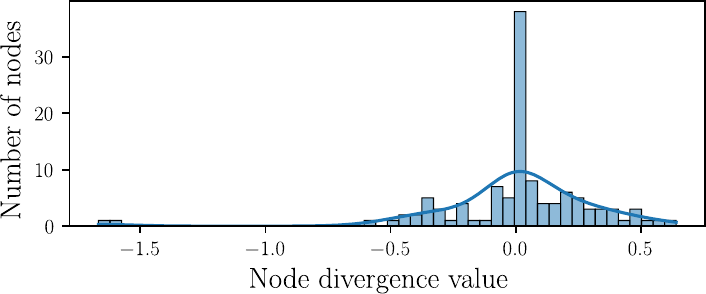}  
        \label{subfig:histogram_div_edge_signals}
    }
    \\
     \subfigure[$\overline{\text{div}}(\mathbf{S}^1)$]
    {
        \includegraphics[width=0.72\columnwidth]{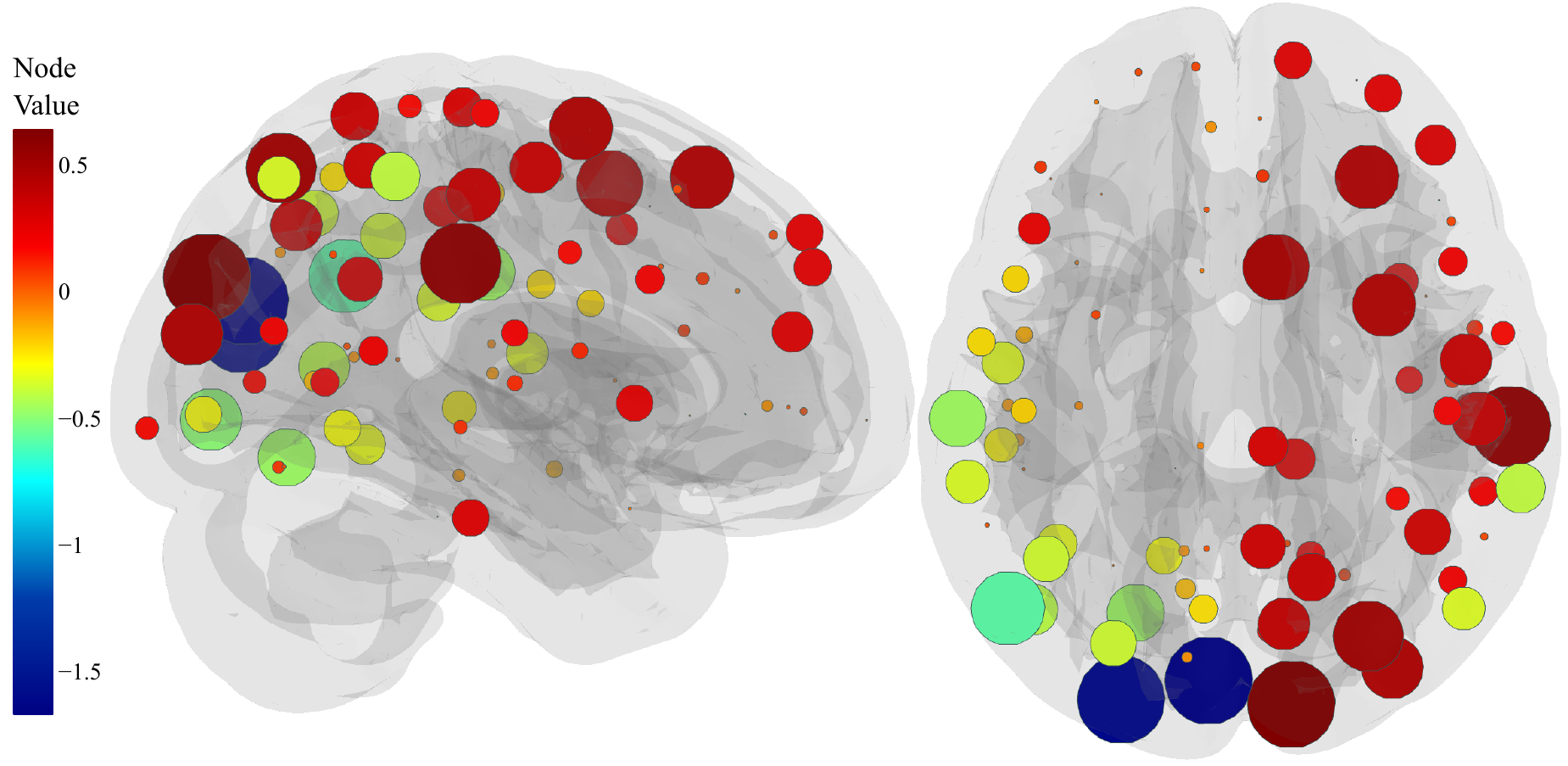}
        \label{subfig:mean_div_edge_signals}
    }
    \\
    \subfigure[Histogram of $\overline{\text{curl}}(\mathbf{S}^1)$]
    {
    \includegraphics[width=6.5cm,height=2.4cm]{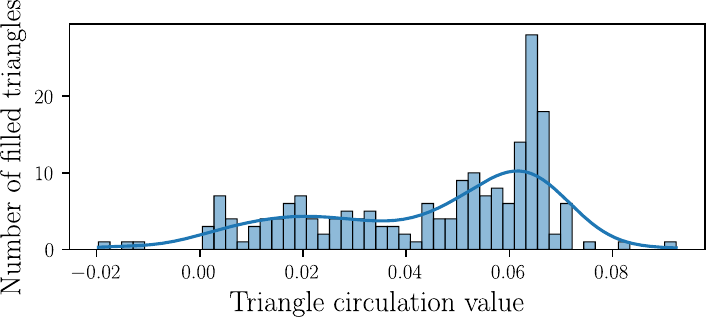}
        \label{subfig:histogram_curl_edge_signals}
    }
    \\
    \subfigure[$\overline{\text{curl}}(\mathbf{S}^1)$]
    {
        \includegraphics[width=0.72\columnwidth]{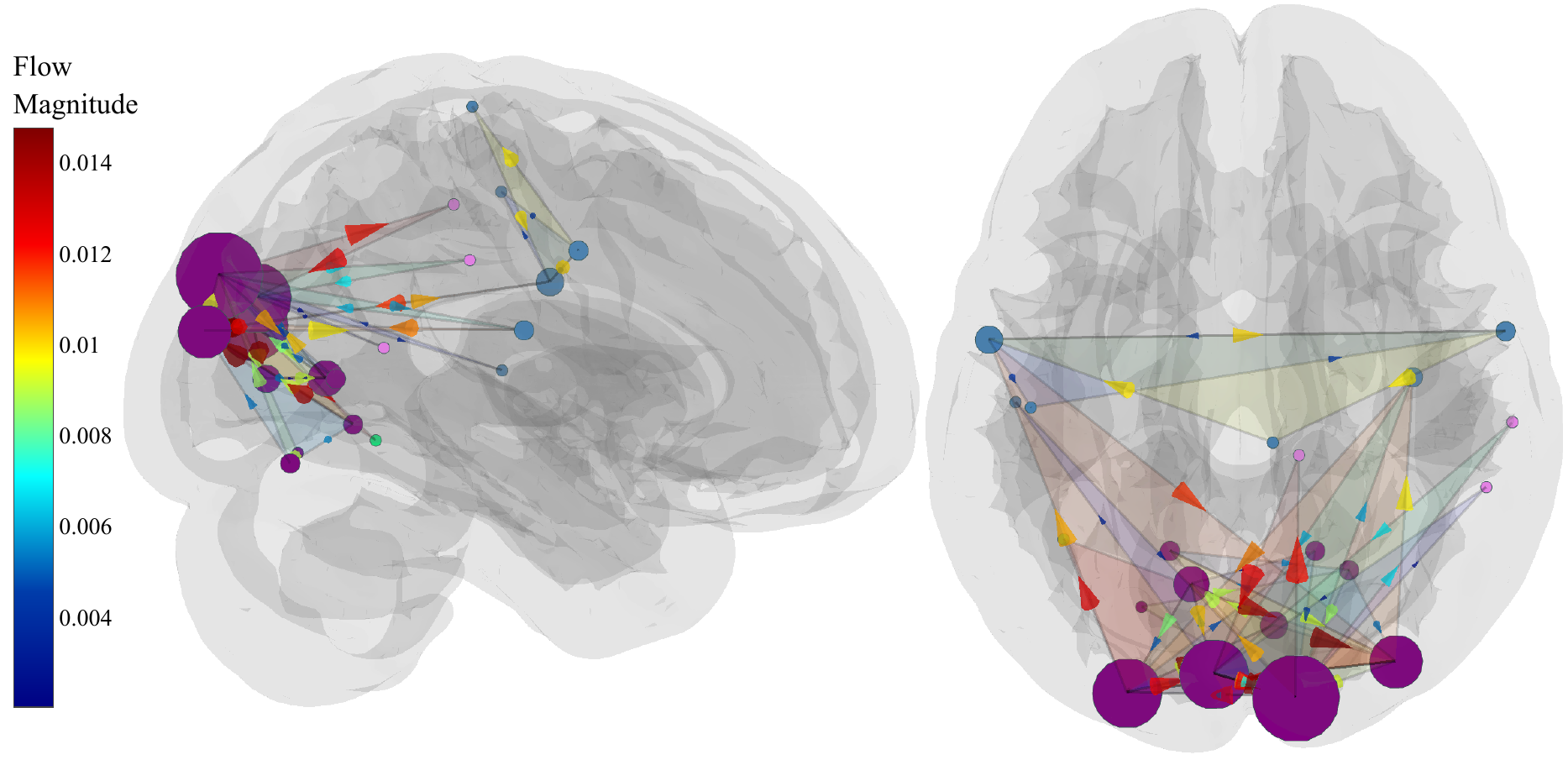}
        \label{subfig:mean_curl_edge_signals}
    }
    \caption{Histograms and 3D brain representations of the averaged divergence and circulation of the edge signals.}
    \label{fig:mean_div_curl_edge_signals}
\end{figure}

Given the brain SCX and edge signals, we perform a comprehensive analysis of averaged edge signal divergence to identify the primary sources and sinks of information throughout the mean individual rs-fMRI recording. For each edge signal $\mathbf{s}^1(m)$ we compute its divergence using \eqref{eq:div_operator} and compute their average across all $M$ time samples given by $\overline{\text{div}}(\mathbf{S}^1)=\sum_{m=1}^{M}\text{div}(\mathbf{s}^1(m))/M$. In Fig.~\ref{subfig:histogram_div_edge_signals} we show a histogram illustrating the distribution of \(\overline{\text{div}}(\mathbf{S}^1)\) across nodes. In Fig.~\ref{subfig:mean_div_edge_signals}, the values of \(\overline{\text{div}}(\mathbf{S}^1)\) are visualized as node signals over the ROIs, where colors indicate their respective values, and node sizes are scaled according to their magnitudes. Analyzing the most negative values of \(\overline{\text{div}}(\mathbf{S}^1)\), 
we can infer that the overall sink of information originates from the left hemisphere of the brain. Specifically, the primary sinks are located in ROIs from the left side of the visual (VIS) subnet, followed by ROIs from default mode network (DMN) and dorsal attention (DA) subnets. On the other hand, exploring the most positive values of \(\overline{\text{div}}(\mathbf{S}^1)\), 
we can infer that the right hemisphere serves as a source of information, with the primary sources located in ROIs from VIS and ventral attention (VA) subnets, followed by ROIs from DA and DMN subnets. These results exhibit patterns similar to recent findings on the characterization of brain wave dynamics~\cite{Roberts2019}.

Furthermore, we conduct an analysis on the mean circulations of flows along the triangles of the inferred brain SCX. Similarly to divergence analysis of the edge signals $\mathbf{s}^1(m)$, we compute their curl counterparts using \eqref{eq:curl_operator} and compute their average across time as $\overline{\text{curl}}(\mathbf{S}^1)=\sum_{m=1}^{M}\text{curl}(\mathbf{s}^1(m))/M$. Fig.~\ref{subfig:histogram_curl_edge_signals} depicts a histogram illustrating the component distribution of \(\overline{\text{curl}}(\mathbf{S}^1)\). 
In this study, we focus on conservative circulations, i.e. the smallest triangle circulation values in magnitude, highlighting its interpretation from a neuroscience point of view. Fig.~\ref{subfig:mean_curl_edge_signals} illustrates the 20 weakest circulation values (in magnitude) over the filled triangles, where the cones along the edges indicate their (counter-)clockwise orientation.
The cone size and color are proportional to the flow magnitude over each triangle. The ROI size is scaled according to the number of participation among the triangles, while its color corresponds to its FS. We can easily infer that ROIs from the VIS subnet (in purple) play a crucial role in the overall context, interacting with ROIs from somatomotor (SM) subnet (in steelblue). From a neuroscience perspective, the circulation patterns involving the VIS and SM subnets simultaneously can be attributed to functional integration, which characterizes global communication between distinct brain modules~\cite{sporns_2013}. Specifically, this kind of integration can be explained by the underlying redundant-dominated structure present in these regions, as highlighted in recent studies~\cite{faes, varley}. On the other hand, circulations that involve solely ROIs from either the SM or VIS subnet reflect the concept of functional segregation, which refers to specialized neuronal processing within distinct modules~\cite{sporns_2013}.


\section{Joint learning of topology and sparse  signals}\label{sec:sparse_learning}

Simplicial complex learning strategies based on edge signal smoothness  have been proposed in \cite{barb_2020},\cite{gurugubelli2024simplicial}. 
In this section, we design a strategy to   jointly learn the simplicial complex topology and  sparse spectral representations of both the solenoidal and harmonic  signals from brain data.
Assuming  the graph topology  given, our goal is to infer 
a topological structure where solenoidal and harmonic signals admit sparse spectral representations,  minimizing jointly the  total signal circulations  and the data-fitting error.
Denoting with $\hat{\ms}^{1}_{s}$, $\hat{\ms}^{1}_{H}$, the SFT coefficients of, respectively, the solenoidal and harmonic signals, their spectral representations can be written as
$\ms^{1}_{s}=\mU_{\text{C}} \,{\hat{\ms}}^1_{s}, \quad \ms^{1}_{H}=\mU_{\text{H}} \,{\hat{\ms}}^1_{H}$.
Given  a set of $M$ edge signals $\mY^1=[\my^{1}(1),\ldots, \my^{1}(M)]\in\mathbb{R}^{E\times M}$,  we need to learn from data the incidence matrix $\mB_2$, or equivalently, the upper Laplacian matrix $\mL_{1,u}$.  Denoting with $T$ the number of possible triangles in the complex, this matrix can be written in the equivalent form $\mL_{1,u}=\sum_{n=1}^{T} q_n \mathbf{b}_n \mathbf{b}_n^T$
where $\mathbf{b}_n$ are the $E\times 1$-dimensional  columns of the matrix $\mB_2$, while $q_n \in\{0,1\}$ are binary  coefficients   selecting the filled triangles according to the chosen optimization rule. Then, similarly to the learning method in \cite{barb_2020}, we first 
check if data have components on the solenoidal subspace 
by projecting the observed edge flow
 signals onto the space orthogonal to the one  spanned by the irrotational component as
 $\my^{1}_{sH}(m)=(\mI-\mU_{\text{G}} \mU_{\text{G}}^T) \my^{1}(m)$, $ \forall m$.
Then, we calculate the Frobenius norm of the matrix $\mY^{1}_{sH}=[\my^{1}_{sH}(1), \ldots, \my^{1}_{sH}(M)]$ 
and if this norm is higher than a given threshold, we proceed to learn the upper Laplacian matrix.
Therefore, defining the matrices $\hat{\mS}_{s}^{1}=[\hat{\ms}^{1}_{s}(1),\ldots, \hat{\ms}^{1}_{s}(M)]$, $\hat{\mS}_{H}^{1}=[\hat{\ms}^{1}_{H}(1),\ldots, \hat{\ms}^{1}_{H}(M)]$, we formulate the following non-convex optimization problem 
\vspace{-0.3 cm}
\beq \label{eq:P_mtv_sign_est1}
\begin{array}{lll}
\underset{ \substack{\hat{\mS}^{1}_{s}, \hat{\mS}^{1}_{H}, \mq \\ \mU_{\text{C}}, \mU_{\text{H}},\boldsymbol{\Lambda}_s}}{\min} & \!\!\!\!\!\ds \sum_{n=1}^{T} q_n \!\parallel \mY^{1\, T}_{sH} \mathbf{b}_n \parallel_F^2\!+\beta \!
\ds  \parallel \mY^1_{sH}{\scriptstyle -}\mU_{\text{C}}\hat{\mS}^{1}_{s}{\scriptstyle -}\mU_{\text{H}}\hat{\mS}^{1}_{H}\parallel_F^2 \\
\quad \text{s.t.} &  a)\, 
\parallel \hat{\ms}^{1}_{s}(m) \parallel_1 \leq \alpha_1, \parallel \hat{\ms}^{1}_{H}(m) \parallel_1\leq \alpha_2, \; \forall m, \medskip\\
 \quad \quad & b)\, \ds \sum_{n=1}^{T} q_n \mathbf{b}_n \mathbf{b}_n^{T} \mU_{\text{C}}= \mU_{\text{C}} \boldsymbol{\Lambda}_s,  \\
\quad \quad & c)\, (\ds \sum_{n=1}^{T} q_n \mathbf{b}_n \mathbf{b}_n^{T}+\mB_1^T \mB_1) \mU_{\text{H}}= \mathbf{0}, \medskip\\
\quad \quad & d)\, \parallel \mq \parallel_0 \leq q^{\star}, \quad \mq \in \{0,1\}^{T},  \quad \quad (\mathcal{P}_t)\\
\end{array}
\eeq
where  the first term in the objective function is the energy of the signals circulation along all triangles, while the second term is the data-fitting error controlled by the positive coefficient $\beta>0$. The $l_1$-norm constraints  $a)$ enforce spectral sparsity of the signals through the positive coefficients $\alpha_1,\alpha_2$, while
$b)$ and $c)$ constrain the matrices $\mU_{\text{C}}$ and $\mU_{\text{H}}$ to be the eigenvectors associated with the non-zeros eigenvalues  of $\mL_{1,u}$, and the eigenvectors spanning the kernel of $\mL_1$, respectively. Finally, constraint $d)$ imposes that the  number of filled triangles doesn't exceed a maximum number $q^{\star}$. Problem in (\ref{eq:P_mtv_sign_est1}) is non-convex due to the constraints $b)$, $c)$ and $d)$,  therefore we propose an alternating iterative suboptimal strategy solving it as a sequence of convex-subproblems. Specifically,  by varying the value  of the required number of filled cells $q^{\star}$ from $1$ to the  maximum value $T$, we alternate between the solutions of two convex optimization problems and then select the number $q^{\star}$ minimizing the data-fitting error. Specifically,  we iterate for $i=1, \ldots, T$ between the following two steps:
\begin{itemize}
    \item[S.1] Solve problem $\mathcal{P}_t$ with respect to $\mq$  with $\beta=0$, $q^{\star}=i$ and considering only constraints $b)$, $c)$ and $d)$. This problem admits the closed form solution obtained by computing $a_n=\parallel \mY^{1\, T}_{sH} \mathbf{b}_n \parallel_F^2$, sorting these coefficients  in increasing order and selecting the indices $\mathcal{T}_t=\{i_1,i_2,\ldots,i_{q^{\star}}\}$ of the $q^{\star}$ lowest coefficients $a_n$. Hence, we get $\hat{\mL}_{1,u}^{i}=\sum_{n\in \mathcal{T}_t} \mathbf{b}_n \mathbf{b}_n^T$;
    \item[S.2] Given $\hat{\mL}_{1,u}^{i}$ find the eigenvectors matrices $\mU_{\text{C}}^{i}$  and, using the  estimated $1$-order Laplacian $\hat{\mL}_{1}^{i}=\mL_{1,\ell}+\hat{\mL}_{1,u}^{i}$, derive the eigenvectors matrix $\mU_{\text{H}}^{i}$  spanning its kernel. Then, solve the following convex optimization problem 
    \beq
\label{eq:P_mtv_sign_est}
\begin{array}{lll} \nonumber
\underset{ \hat{\mS}^{1}_{s}, \hat{\mS}^{1}_{H}}{\min} & \ds\parallel \mY^1_{sH}{\scriptstyle -}\mU_{\text{C}}^{i}\hat{\mS}^{1}_{s}{\scriptstyle -}\mU_{\text{H}}^{i}\hat{\mS}^{1}_{\text{H}}\parallel_F^2 \quad \quad (\mathcal{P}_{i})\\
\quad \text{s.t.} &  a)\, 
\parallel \hat{\ms}^{1}_{s}(m) \parallel_1 \leq \alpha_1, \;  \parallel \hat{\ms}^{1}_{H}(m) \parallel_1\leq \alpha_2, \; \forall m. \medskip\\
\end{array}
\eeq     \vspace{-0.2cm} 
\end{itemize}
In each iteration $i$, we find the optimal data fit error $\text{g}(i)=\ds\parallel \mY^1_{sH}{\scriptstyle -}\mU_{\text{C}}^{i}\hat{\mS}^{1}_{s}{\scriptstyle -}\mU_{\text{H}}^{i}\hat{\mS}^{1}_{\text{H}}\parallel_F^2$ by alternating between steps $\text{S.1}$ and $\text{S.2}$. Finally, we derive the filled triangles as  $q^{\star}= \arg\min_{i \in \{1,\ldots, T\}} \text{g}(i)$ and the learned Laplacian $\hat{\mL}_{1}=\mL_{1,\ell}+\hat{\mL}_{1,u}^{q^{\star}}$.

\begin{figure}[!t]
    \centering

    \subfigure[Histogram of $\overline{\text{curl}}(\mathbf{S}^1)$]
    {
     \includegraphics[width=6.5cm,height=2.4cm]
{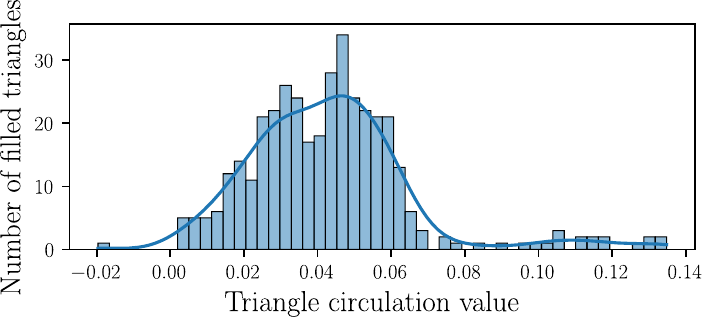}     
        \label{subfig:histogram_curl_edge_signals_sparse}
    }
    \\
    \subfigure[$\overline{\text{curl}}(\mathbf{S}^1)$]
    {
        \includegraphics[width=0.72\columnwidth]{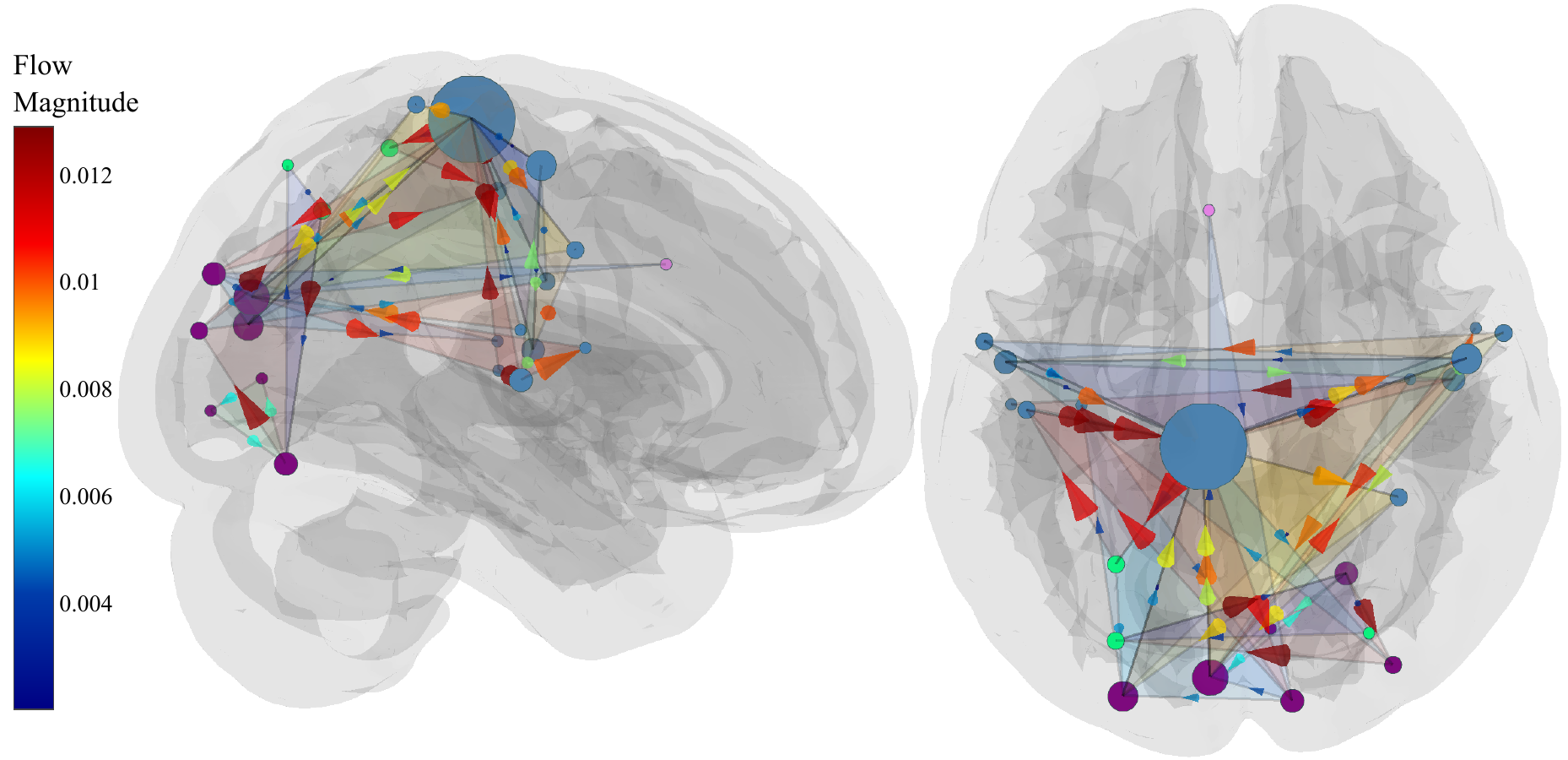}
        \label{subfig:mean_curl_edge_signals_sparse}
    }

    \caption{Histograms and 3D brain representations of the averaged divergence and circulation of the  signals estimated using the joint learning strategy.}
    \label{fig:mean_div_curl_edge_signals_sparse}
\end{figure}

As numerical test, we apply the proposed iterative algorithm to learn  the brain topology using the same brain graph and edge signals $\mathbf{S}^1$ from Sec.~\ref{sec:statistical_learning}. Hence, we get  a mean brain SCX with \( T = 381 \) filled triangles.
Subsequently, we analyze the flow circulations along the filled triangles while the divergence patterns persist, as shown in Fig.~\ref{subfig:mean_div_edge_signals}. Fig.~\ref{subfig:histogram_curl_edge_signals_sparse} shows the component distribution of \(\overline{\text{curl}}(\mathbf{S}^1)\), while Fig.~\ref{subfig:mean_curl_edge_signals_sparse} highlights its primary conservative circulations. Similar patterns to Fig.~\ref{subfig:mean_curl_edge_signals} are observed, including circulations involving both intra- and inter-module HOIs among ROIs from SM and VIS subnets. This reinforces the concepts of functional segregation~\cite{sporns_2013} and integration, driven by the redundant-dominated structure shared between these subnets~\cite{faes, varley}.



\vspace{-0.1 cm}
\section{Conclusion}\label{sec:conclusion}

In this paper we characterized HOIs in the brain for the first time  through the lens of the TSP framework. We developed two approaches for learning the mean brain topology from real brain datasets using higher-order statistical measures and TSP tools. 
By analyzing the divergence and circulation properties of brain edge signals, 
our study reveals that brain conservative signals, i.e., circulation signals with smallest magnitudes, reflect concepts of functional segregation and integration. For instance, both learning approaches reveal functional integration between the SM and VIS subnets, consistent with recent neuroscience findings on redundant-dominated structures in the brain. 


\bibliographystyle{IEEEbib}
\bibliography{reference}

\end{document}